# Surface plasmon resonance biosensors based on Kretschmann configuration: basic instrumentation and applications

Nishant Shukla[1], Pawan Chetri[2], Ratan Boruah[1], Ankur Gogoi[3,*], Gazi A Ahmed[1,*]
[1]Department of Physics, Tezpur University, Tezpur 784028, Assam, India
[2]Department of Physics, D. R. College, Golaghat 785621, Assam, India
[3]Department of Physics, Jagannath Barooah College, Jorhat 785001, Assam, India
*Corresponding author: ankurgogoi@gmail.com; gazi@tezu.ernet.in

**Abstract**

Surface Plasmon Resonance (SPR) offers a powerful tool for label-free and non-invasive characterization of biomolecular interactions. To date, several experimental configurations, based on two fundamental physical phenomena, e.g., attenuated total reflection and diffraction, have been developed to measure the SPR signal generated due to the resonant interactions between incident light and plasma waves on the metal surface. These configurations are divided into three categories: grating-based, prism-based, and waveguide-based coupling. Among such techniques, one of the prism-based SPR coupling schemes, popularly known as Kretschmann configuration, is most widely used due to its high sensitivity, operational simplicity, lower cost, and real-time detection. This chapter explains the basic instrumentation and reviews the recent trends in the development of Kretschmann configuration-based SPR biosensors with its applications.

## 1. Introduction

The great Czech medical scientist Johannes Purkinje coined the term "Plasma" for the clear liquid that remains after removing various corpuscles of the blood. In 1927 the term was borrowed by the Noble prize-winning American chemist Irving Langmuir in order to describe the electrified fluid-carrying electrons and ions as an analogy to blood plasma carrying red and white corpuscles [1]. During the investigation related to the extension of the lifetime of a filament [2] he developed the theory of plasma sheaths and Langmuir waves- the periodic variations of the electron density [3]. In 1932, Steenbeck, applied the theory of Tonks and Langmuir, suggesting the existence of plasma oscillations in metals [4]. The free conduction electrons in the metal interact with electromagnetic radiation and can sustain surface and volume charge density oscillations called plasmons distinct resonance frequencies. The first such behaviour was observed in 1902 by Wood, who suggested implementing the idea of polarization to explain the singular behaviour of grating [5]. Later, the different roles of surface and their importance were envisioned in the work of Zenneck (in 1907) [6] followed by Sommerfeld (in 1909) [7], Fano (in 1941) [8] and Bouwkamp (in 1950) [9], who analysed the solution of Maxwell's equations that had "surface-wave" (plasmons) properties. The subject gained its realization after a series of publications by Bohm and Pines during 1950s [10]. In 1968, Otto manifested the coupling of light waves to surface plasmon waves by using the method of frustrated total reflection [11]. In the same year, Kretschmann and Raether, through their practically developed method based on total internal reflection advanced for the advertised use of SPR [12].

SPR technique relies on measurement of the change in the refractive index of the sensing layer due to molecular interactions on the sensor surface [13]. This technique relies on the certitude that, for certain defined conditions, surface plasmons on a metallic film can be excited by photons at the interface of two materials [14]. The certainty of the condition relies on the analysis of a metal/dielectric interface by Maxwell's equations. Although the history of biosensors goes back to 1962 with Leland C. Clark developing the enzyme electrodes [15], it was in 1983 that Liedberg, Nylands and Lundstroem for the first time demonstrated the physical methods for the label-free, real-time detection of biomolecules using SPR technique [16-18]. Basically, biosensors are devices that analytically converts a biological response into an electrical signal [19]. In SPR sensors, the extent of coupling between the incident wave and the surface plasmons are elucidated through angular, wavelength, intensity, phase, and/or polarization modulation schemes by measuring the parameters that yield the strongest coupling, which is subsequently used as the sensor output (calibrated to refractive index) [20]. Since its first attempt in the field of biosensors by Liedberg et al. [17] in 1983 and the induction of the first commercial SPR instrument (BIAcore) by Pharmacia Biosensor in 1990 [21,22], SPR sensing has become a gold standard for the measurement of various biomolecular interactions [23]. Starting from the study of binding kinetics and affinity of protein-protein or protein-lipid interactions to viral diagnostics, the profound success of SPR has been documented by more than 30,000 research works, registered in PubMed, since 1980s.

Notably, among various SPR sensor types, Kretschmann configuration-based sensors are the most widely used biosensor due to their inherent advantages, such as high sensitivity, operational simplicity, and real-time detection capability. In this chapter, a brief discussion on the instrumentation and various applications of Kretschmann configuration based SPR biosensors is presented.

## 2. Theoretical background

The collective charge (electron) density oscillation occurring on a metal surface is called plasma oscillation and the quantum associated with this oscillation is known as surface plasmon (SP) [24]. SPs can be excited by electrons [25,26], phonons [27] and photons [28,29]. Under specific conditions, such as, when light of suitable wavelength is shone at the interface between a semi-infinite metal layer and a dielectric medium, the phase matching conditions between the optical and surface plasma wave (SPW) vector are achieved and SPs are excited. Such a condition is known as surface plasmon resonance (SPR) and the resultant surface trapped electromagnetic oscillation due to the resonant coupled interaction between the incident light and SPW is defined as a surface plasmon polariton (SPP) [30-33]. The excitation of SPs in case of metallic nanostructures is known as localized surface plasmon resonance (LSPR).

The wavevector associated with a SPW at a metal-dielectric interface is given by [29,34-36]:

$$k_{spp} = k_0 \sqrt{\frac{\epsilon_m \epsilon_d}{\epsilon_m + \epsilon_d}} \qquad (1)$$

where, $k_0 = \frac{\omega}{c}$ is the wavevector of the incident wave, $\omega$ is the incident frequency, $c$ is the speed of light, $\epsilon_m$ and $\epsilon_d$ are the metal layer permittivity and dielectric (analyte) permittivity, respectively. Notably, the dielectric constant of the metal is a complex quantity, i.e., $\epsilon_m = \epsilon_m' + i\epsilon_m''$, where $\epsilon_m'$ and $\epsilon_m''$ represents the real and imaginary part, respectively. Thus, the wave vector of the SPW is also a complex quantity, i.e., $k_{spp} = k_{spp}' + ik_{spp}''$ with [29]

$$k_{spp}' = \frac{\omega}{c} \left(\frac{\epsilon_m' \epsilon_d}{\epsilon_m' + \epsilon_d}\right)^{\frac{1}{2}} = \frac{2\pi}{\lambda} \left(\frac{\epsilon_m' n_d^2}{\epsilon_m' + n_d^2}\right)^{\frac{1}{2}} \qquad (2)$$

$$k_{spp}'' = \frac{\omega}{c} \left(\frac{\epsilon_m' \epsilon_d}{\epsilon_m' + \epsilon_d}\right)^{\frac{3}{2}} \frac{\epsilon_m''}{2(\epsilon_m')^2} = \frac{2\pi}{\lambda} \left(\frac{\epsilon_m' n_d^2}{\epsilon_m' + n_d^2}\right)^{\frac{3}{2}} \frac{\epsilon_m''}{2(\epsilon_m')^2} \qquad (3)$$

where, $\epsilon_d = n_d^2$ [37], $n_d$ being the refractive index of the dielectric (analyte). As can be seen from equation (2), for real values of $k_{spp}'$, one must have $\epsilon_m' < 0$ and $|\epsilon_m'| > \epsilon_d$. Notably metals, such as, gold and silver, satisfy this condition[38]-[40].. Moreover, $k_{spp}''$ determines internal absorption, which is responsible for the dissipation of the energy of the SPs on smooth surfaces [29].

On the other hand, merely shining light on a smooth metal surface may not excite surface plasmons since the wavevector of photons in free space, $k_0$, is smaller than that of SP [40,41], as shown in the dispersion relation curve (Fig. 1). Thus, there is always a mismatch between the momenta of the incident light photon ($\hbar k$) and SP ($\hbar k_{spp}$), for the same frequency. Nevertheless, several momentum-compensation coupling mechanisms have been developed, as described in the next section, to increase the wave vector of incident light in the dielectric medium (given by equation 4) to achieve the matching conditions for the excitation of SPs.

$$k = k_0 \sqrt{\epsilon_d} = \frac{2\pi}{\lambda} n_d \qquad (4)$$

In the case of Kretschmann configuration, the wave vector of the incident light is modified as [40]

$$k_x = \frac{2\pi}{\lambda} n_p \sin\theta \qquad (5)$$

where $n_p$ and $\theta$ are the refractive index and angle of incidence in the prism, respectively.

At resonance condition [42,43]

$$k_x = \frac{2\pi}{\lambda} n_p \sin\theta_r = \frac{2\pi}{\lambda} \left(\frac{\epsilon_m' n_d^2}{\epsilon_m' + n_d^2}\right)^{\frac{1}{2}} \qquad (6)$$

$$\theta_r = \sin^{-1}\left\{\left(\frac{1}{n_p}\right)\left(\frac{\epsilon_m' n_d^2}{\epsilon_m' + n_d^2}\right)^{\frac{1}{2}}\right\} \qquad (7)$$

where, $\theta_r$ = coupling angle (i.e., angle at which resonance condition is satisfied)

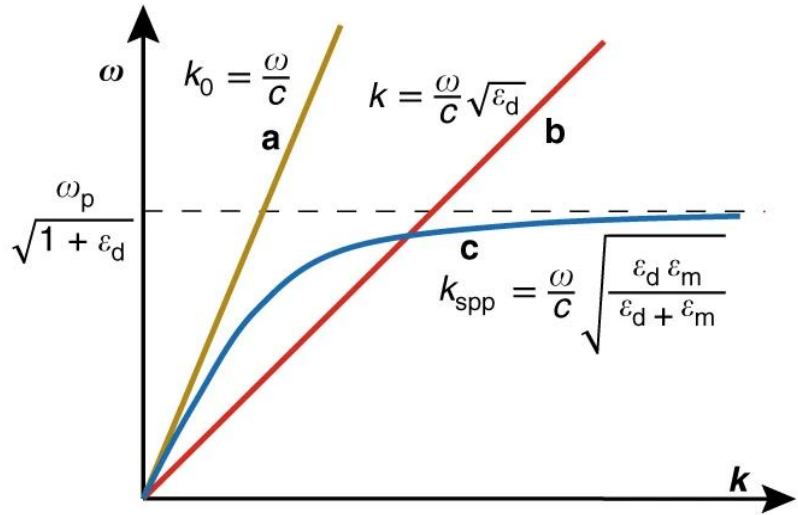

Fig. 1. Graphical representation of the dispersion relation of light in (a) free space, (b) dielectric medium and (c) surface plasmon polaritons. Here, $\omega_P$ is the frequency of bulk longitudinal electron excitations, i.e., the plasma frequency. The wave number, $k$, of a photon is always less than that of SPP at all frequencies. Excitation of SPs can be achieved by the judicious use of dielectric materials that increases the light wave vector so that the dispersion lines intersect. The figure is reproduced from Zhang et al. [34] Copyright © The Author(s) 2021, Creative Commons Attribution 4.0 International License.

## 3. SPR sensor configurations

### 3.1 Experimental principle

At SPR, most of the incident light energy is resonantly coupled into the SPP mode, resulting in a dip in the observed reflectance profile. This resonance condition is localised at the metal-dielectric interface and susceptible to any change in the optical properties (e.g., refractive index) of the dielectric triggered by the binding interactions between sensor-immobilized ligands and sample biomolecules (analyte), and therefore can be exploited as a method for detecting external medium near the interface (Fig. 2A) [44-48]. For instance, when a biomolecular binding event occurs on a gold surface, the mass of the accumulated biomolecules increases. While this leads to an increase in the refractive index of the sensing surface, the prism refractive index remains the same [49], resulting in significant changes in the resonance angle, wavelength, intensity and/or phase of the reflected light which is typically detected by measuring the variation in the reflected light (Fig. 2B). In addition, time rate of change of the resonance or response units (RU), that signifies the increase or decrease in the signal, can also be measured in the form of a sensorgram, as shown in Figure 2C. Such capabilities of SPR biosensors allow for real time quantification of the amount of surface concentration of the analyte molecules and binding affinity, and thus, label-free monitoring of biomolecular reaction kinetics. Typically, 1000 RU represents an angular shift of 0.1° [50-52]. A SPR biosensor is schematically represented in Fig. 2.

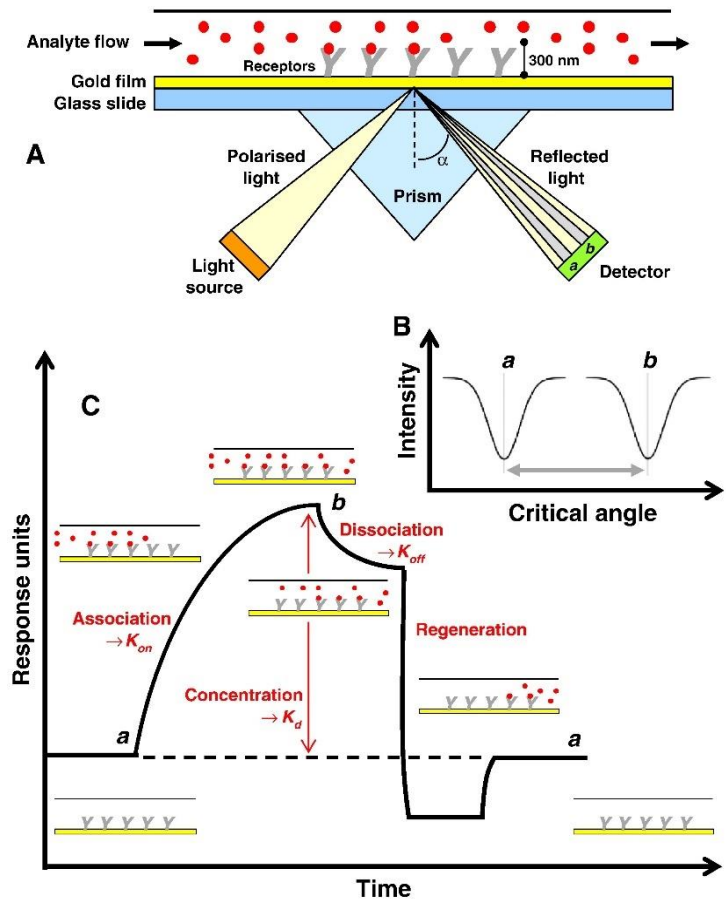

Fig. 2. Schematic representation of SPR biosensing principle. A. Prism-coupled configuration based SPR instrument set up. B. Shift of the resonance angle from *a* to *b* when analyte molecules bind to receptor molecules. C. Four phases of a typical SPR sensorgram: association phase, equilibrium phase or steady state, dissociation phase, and regeneration phase. $K_{on}$: rate of association, $K_D$: binding affinity, $K_{off}$: rate of dissociation. The figure is reproduced with permission from Patching [52], Copyright © 2013 Elsevier B.V.

### 3.2 Basic Instrumentation

The keystone in the construction of a SPR sensor is the efficient coupling between the incident light wave and the SPs at the metal-dielectric interface. Since SPW is a TM wave (p-polarized wave, i.e., the electric field vector is parallel to the incident plane), SPR is possible only when the wavevector of the incident light parallel to the interface (i.e., TM or p-polarized light) becomes equal to that of the SPW. Notably, the wavevector of incident (p-polarized) light can be increased and matched with that of SPW to facilitate efficient coupling by using specially designed optical coupling configurations. There are three coupling configurations that can be used to optically excite surface plasmons: grating-based, prism-based, and optical waveguide-based coupling (Fig. 3) [24].

In a grating coupler-based configuration, first demonstrated by Wood [5,53] in 1902, the resonance condition is achieved by diffraction of the incident light. Briefly, a light beam incident on a periodically distorted metal-dielectric interface (like a grating) gets diffracted in different directions with an increase or decrease of the incident light wave vector by integer multiples of the grating wave vector $k_g$ ($k_g = 2\pi/\lambda$). A diffracted beam becomes evanescent when the diffracted order has a wave vector greater than that of the incident grazing radiation. SPR is achieved when the wavevector of the parallel component of the diffracted light matches that of the SPs [48].

Otto [11], Kretschmann and Raether [12] introduced the excitation of SP by using prism coupling in 1968. Otto configuration is based on the principle of total internal reflection occurring at the prism/dielectric (air) interface and subsequent penetration of an evanescent field into the outer layer. When a nano-dimensional metal film is brought close to the outer layer of the prism, the electromagnetic field of the SP at the metal/dielectric interface couples to the incident evanescent wave provided the resonance condition is reached [11]. This technique is called the attenuated total internal reflection (ATR) technique since the evanescent wave is generated by the phenomenon of total internal reflection in the prism and subsequently coupled with the SPs. On a similar basis, Kretschmann and Raether configuration (popularly known as the Kretschmann configuration) relies on ATR. Nevertheless, in contrast to Otto configuration, the Kretschmann configuration involves the deposition of a nano-dimensional metal thin-film directly on the hypotenuse surface of a right-angle prism. The evanescent fields produced by total internal reflection pass through the nano-dimensional metal film and, if the thickness of the metal film is less than the skin depth of the material, the SP can easily couple to the outer layer. Notably, optimum coupling at visible light wavelengths can be achieved for a metal film thickness in the range of 10 nm to 70 nm [54].

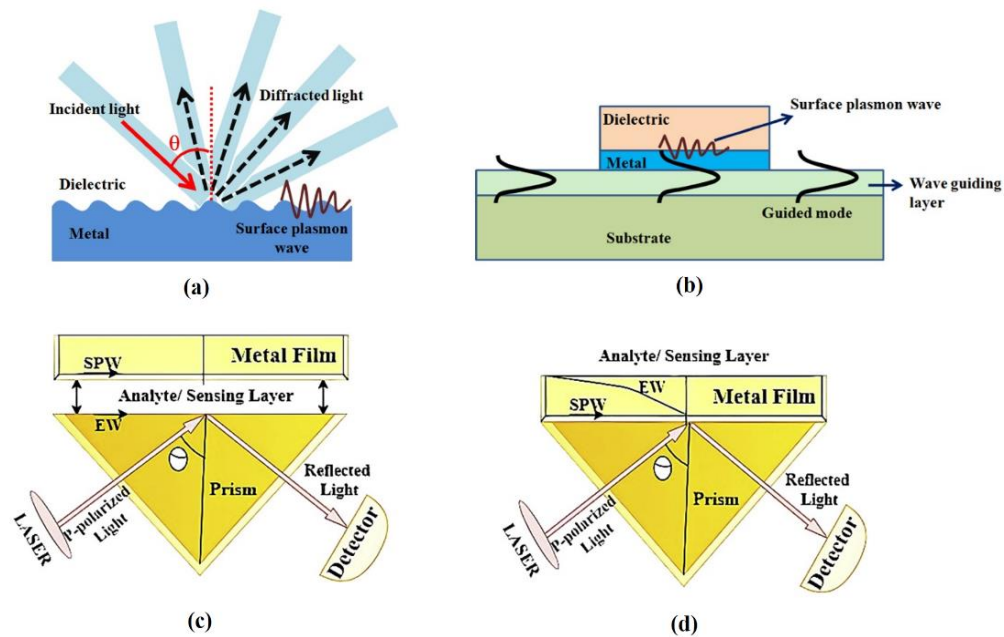

Fig. 3. Basic configurations of SPR sensors: (a) grating-coupled SPR system; (b) optical waveguide-coupled SPR system; prism-coupled (c) Otto configuration and (d) Kretschmann configuration. Figures (a) and (b) are reproduced from Gupta et al. [55], Copyright © 2016 by the authors; licensee MDPI, Basel, Switzerland, Creative Commons Attribution (CC-BY) license. Figures (c) and (d) are reproduced from Yesudasu et al. [39], Copyright © 2021 Elsevier Ltd., Creative Commons Attribution (CC BY-NC-ND) license.

Similarly, an optical waveguide-coupled SPR configuration is also based on ATR. The waveguide guides a light wave to the region deposited with a metallic thin film. If the surface plasma wave (SPW) and the guided light wave that evanescently penetrate the metal layer are phase-matched, the light wave excites the SPs at the outer interface of the metal film [48].

Table 1. Comparison of different SPR coupling configurations

| SPR coupling configuration | Advantages | Disadvantages |
|---|---|---|
| Grating-based | Higher integration capabilities, possibility if miniaturized sensor fabrication for lab-on-a-chip applications [56] | sensitivity is comparatively low; complicated manufacturing process [38,57] |
| Prism-based | Highly sensitive, easy to use [56] | complicated opto-mechatronic instrumentation, especially for angular modulation; miniaturization is usually difficult since the size of the prism is large [556,58,59] |
| Waveguide-based | Miniaturized sensor size [58], highly sensitive, no moving parts [44] | angular modulation is not applicable, SPR reflectivity profile can be obtained with polychromatic light only [43] |

Although all these configurations have their unique advantages and limitations, as summarized in Table 1, the prism-based coupling is the most widely used method, especially for biosensing, due to its high sensitivity and operational simplicity [56]. Notably, between the two prism coupling strategies, Kretschmann configuration is the most conventional approach for the fabrication of SPR based sensors [60], since more efficient plasmon generation is possible in this configuration due to the presence of the metal layer directly on top of the prism surface [31]. In contrast, Otto configuration requires the metallic layer to be brought within a few microns of the prism and maintain a constant air gap over the entire surface, which is very difficult to accomplish [61]. But it sometimes can happen that the layer required over the substrate does not stick due to its poor ability of sticking, requiring adhesion. The Otto configuration seems to be more efficient there as the dielectric and refractive index properties of substrate will now have extra dependent parameters related to this adhesion to be taken care of. [Barchiesi, D., Grosges, T., Colas, F., & de la Chapelle, M. L. (2015). Combined SPR and SERS: Otto and Kretschmann configurations. *Journal of Optics*, *17*(11), 114009.]

### 3.2 SPR measurement methodologies

Notably, biochemical interactions at the sensor surface trigger local changes in the refractive index of the dielectric near the metal layer, leading to alterations in the propagation constant of the SPs. Consequently, the coupling condition between the light wave and SPs gets altered, which affects the characteristic properties of the light wave after interacting with the SPs. Notably, at SPR, both amplitude and phase of the p-polarized light changes significantly due to the momentum transfer between light and SPs, while those of the s-polarized light is changed slightly [62]. Based on these changes and depending on the intended characteristics of the light wave to be measured, SPR sensors can be classified into angular, wavelength, intensity, phase, and/or polarization modulation sensors [14,43,63,64], as shown in Fig. 4.

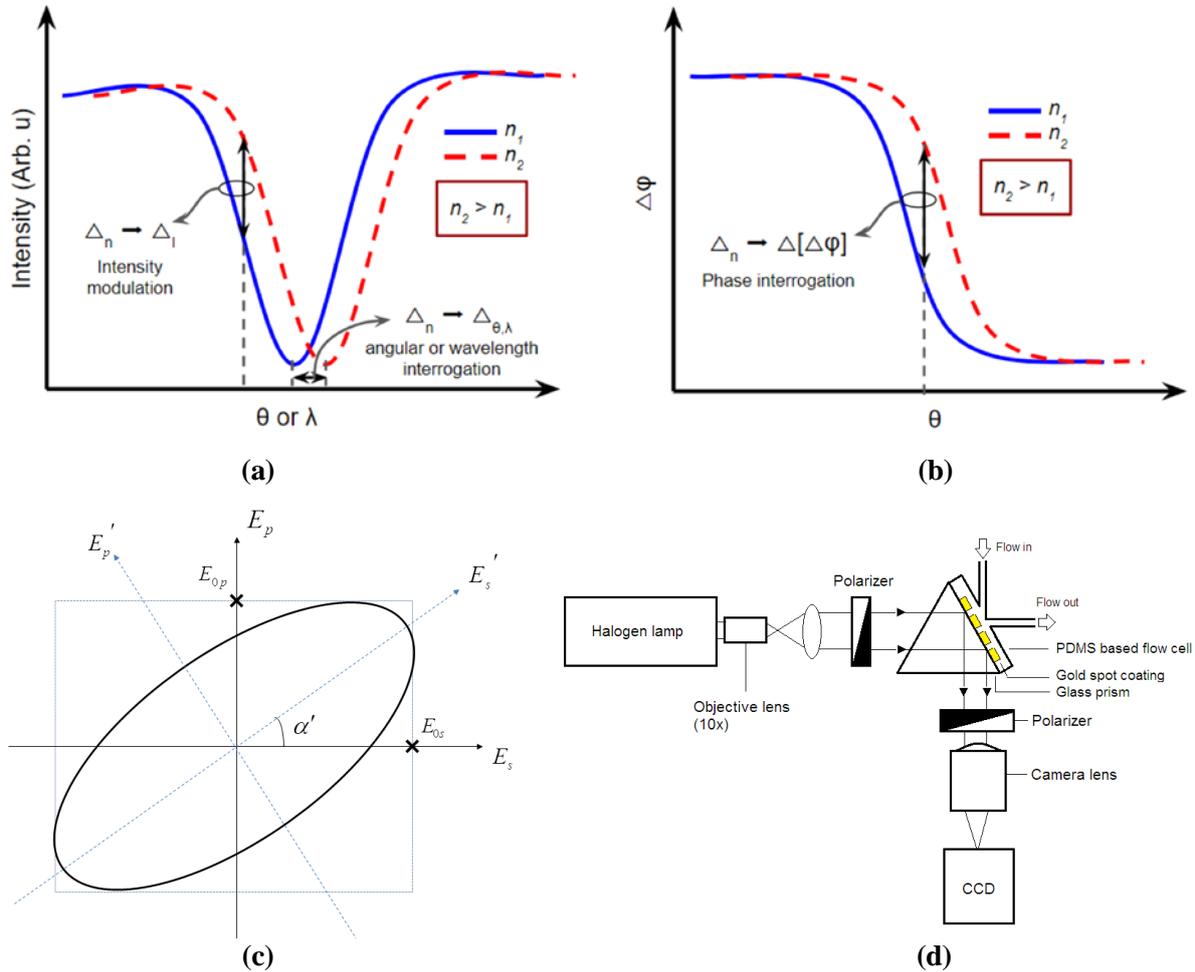

Fig. 4. SPR measurement methodologies: (**a**) angular, wavelength, intensity modulation, and (**b**) phase modulation method for SPR measurements. The figure is reproduced from Prabowo et al. [43], Copyright © 2018 by the authors, Creative Commons Attribution (CC BY) license. (c) The phase difference $\Delta\varphi$ between the p- and s- polarization components produce a shift $\alpha'$ in the orientation angle of the light ellipse, (d) Schematic representation of the SPR experimental setup used for polarization modulation SPR measurement method. The figure is reproduced with permission from Wong et al. [65], Copyright © 2013 Elsevier B.V.

In angular modulation, a monochromatic light source is used and the intensity of the reflected light is found out by scanning throughout the angular range. Likewise, a broadband light source is used in wavelength modulation and the reflected light intensity is measured by varying the wavelength at a fixed incident angle. Thus, if the output signal in the form of normalized reflected intensity (R) is measured as a function of incident angle (θ) or wavelength (λ), by keeping other parameters and components (e.g., the thickness of the metal and dielectric layers) unaffected, then a sharp dip is observed at resonance angle $\theta_{sp}$ of wavelength $\lambda_{sp}$ due to an efficient transfer of incident photon energy to SPs. On the other hand, intensity modulation is primarily used for SPR imaging and is based on the detection of reflectance intensity variation at a fixed incident angle or wavelength

[64,66,67]. As the name suggests, the phase interrogation method involves the detection of the phase shift in the reflected optical beam with respect to the incident beam [68,69] and the polarization modulated SPR sensors exploit the polarization ellipse orientation caused by the phase difference between the p-polarization and s-polarization components [63,70].

### 3.3 Performance Parameters

The performance of any SPR sensor can be defined by various parameters e.g., sensitivity, resolution, detection limit, linearity, dynamic range. Although these parameters have been covered by several excellent publications [39,44, 49,71-75], the major performance parameters (viz., sensitivity, resolution and limit of detection) in case of a Kretschmann configuration based (prism-coupled) SPR sensor operated in angular and wavelength modulation modes are briefly described here, in this section.

Sensitivity of a SPR sensor signifies the detectable fluctuations in the magnitudes of the SPR parameters of interest (e.g., resonant angle, wavelength, intensity) with a change in refractive index of the sensing layer. Larger the shift, higher is the sensitivity. For a Kretschmann configuration based SPR sensor, the sensitivities for angular modulation and wavelength modulation can be derived from equation (6) and are given by [31,42,72,76],

$$S_{p,\theta} = \frac{d\theta}{dn_d} = \frac{\epsilon'_m \sqrt{-\epsilon'_m}}{(\epsilon'_m + n_d^2)\sqrt{\epsilon'_m(n_d^2 - n_p^2) - n_d^2 n_p^2}} \tag{8}$$

$$S_{p,\lambda} = \frac{d\lambda}{dn_d} = \frac{(\epsilon'_m)^2}{\frac{n_d^2}{2}\left|\frac{d\epsilon'_m}{d\lambda}\right| + (\epsilon'_m + n_d^2)\epsilon'_m\left|\frac{dn_p}{d\lambda}\right|\frac{n_d}{n_p}} \tag{9}$$

Resolution is the minimum change in the SPR parameter (e.g., refractive index) that can be resolved by the sensor. It depends on the standard deviation of the output noise ($\sigma_{so}$) and bulk refractive index sensitivity ($S_{RI}$) of the sensor, and is expressed as [72,73,77],

$$\sigma_{RI} = \frac{\sigma_{so}}{S_{RI}} \tag{10}$$

where,

$S_{RI} = \left|\frac{\delta\theta_r}{\delta n_{eff}}\right|\frac{\delta n_{eff}}{\delta n}$ (for angular modulation)

$= \frac{\delta n_{eff}}{\delta n}$ (for wavelength modulation)

$\lambda_r$ = coupling wavelength

$n_{eff} = \frac{k_{spp}}{k_0}$

$\frac{\delta n_{eff}}{\delta n} = \left(\frac{\epsilon'_m}{\epsilon'_m + n_d^2}\right)^{\frac{3}{2}}$ = sensitivity of the effective index of SP to a bulk refractive change.

Finally, the limit of detection (LOD) or detection limit, is defined as the minimum concentration of the analyte that produces the smallest resolvable change in the measured SPR parameter with reasonable certainty and reproducibility [78-80].

## 4. Applications of Kretschmann configuration based SPR biosensors

SPR sensors have proven to be a highly efficient tool for label-free, non-invasive, real-time monitoring of biomolecular interactions [81]. Recent years have witnessed tremendous advancements in SPR biosensors [82-85] that have already found widespread applications in diverse fields including detection and characterizations of bacteria and viruses [86-88], analysis of biomolecules [89,90] (e.g., protein-nucleic acid [91,92], nuclear receptor- DNA [93] , DNA- DNA interactions [94]), immune recognition [95], drug screening [52], etc. In this section, although a comprehensive review of the applications of SPR biosensors based on Kretschmann configuration will not be presented, as they have been covered by many other works [52,96,97], a brief review of the most recent applications of Kretschmann configuration based SPR biosensors on the following important domains will be discussed. In addition, a representative list of different applications of Kretschmann configuration based SPR biosensors using different modulation techniques are presented in Table 2.

### 4.1 Analysis of lipid/Protein molecules

Lipids and proteins are the key building blocks of life. In addition to forming cell membranes, storing up energy, acting as chemical messengers, maintaining temperature, etc., these molecules play vital roles in the structure and function of living cells. For instance, whereas the structure of cellular membranes is composed of lipids, their specific functions are mainly controlled by proteins. Apart from participating in the distribution and localization of membrane proteins and thereby regulating the cellular signalling and trafficking processes, many membrane lipids act as ligands for different signalling proteins and enzymes [98]. On the other hand, membrane proteins account for approximately 30% of the human genome and constitute more than half of the current drug targets. They also take part in many physiological functions, e.g., molecular recognition, cell adhesion, etc. [99-103]. Therefore, protein-protein, protein–lipid and lipid–lipid interactions are important not only to understand the science behind the physiology of cells, but also for the validation and discovery of newer drugs.

Notably, most of the traditional methods used to study the kinetics of biomolecular interactions, e.g., ELISA (Enzyme-Linked Immunosorbent Assay), MicroScale Thermophoresis (MST), Equilibrium Dialysis, and Affinity Chromatography, require labelling and are therefore complicated and time consuming [104,105]. In this context, SPR based biosensing has found widespread application in the study of real time binding kinetics, affinity, concentration, etc., without labeling the ligand or the analyte, thereby keeping the molecular properties unchanged. Through its technique of immobilizing a ligand on a surface with monitored interaction of analytes, SPR is successfully used for the study of various protein/lipid mechanisms with high sensitivity and high throughput of samples [97,106-111].

Kretschmann configuration based SPR biosensors have been used extensively for the study of the interaction of proteins and/or lipids with each other and with other biological macromolecules [112-116]. Many of these works used BIAcore™ technology which is one of the pioneering commercial instruments for SPR biosensing [52,96,114,117,118]. The standard BIAcore™ setup for SPR experiments is based on prism-coupled Kretschmann configuration and is shown in Fig. 2(a). Besides, these biomolecular interactions are typically studied at specially prepared surfaces known as 'sensor chips' that enable the attachment of analytes on the surface-bound ligands [119-123]. A large number of commercial sensor chips have been developed till date [97,109,124,125]. Notably, the sensor chips HPA (hydrophobic association analysis) and L1 are the most widely used chips developed by BIAcore™ to study protein–membrane interactions [96,108]. Similarly, CM5 is another all-purpose sensor chip from BIAcore™ where different types of ligands including proteins, nucleic acids, carbohydrates, etc., can be attached by using covalent coupling technique [126,127]. In one of such works, Currie et al. [121] created lipid monolayer on a chip by injecting extruded lipid vesicles to measure the binding kinetics of 100 nM PDK1 (3-Phosphoinositide-dependent protein kinase-1) and PKB (Protein kinase B). The kinetics then help in the identification of various PtdIns whose presence make the protein kinase phosphorylate. Notably, phosphoinositide are acidic phospholipids in cell membranes with their principal role in the activation, suppression, regulation of the integral protein activities. Similarly, Bahloul et al. [128] used CM5 sensor chips to immobilize harmonin-a and CDH23 (cadherin-23) to study harmonin-a/myosin VIIa and cadherin-23/myosin VIIa interactions by allowing myosin VIIa tail to flow over the chip at a rate of 30 μl/min for 120 s. In the same work, the group used lipophilic L1 chip to capture large unilamellar vesicles (LUVs) and monitor protein–lipid interactions by allowing proteins to flow over the LUVs at a rate of 10 μl/min for 300 s.

In addition to the commercially available chips, development of other sensor chips have also been reported [39,43,130]. Recently, Vala et al. [131] used a novel plasmonic chip prepared by depositing 1.5 nm Ti adhesion layer and 50 nm gold layer on a glass (BK7) substrate. On the top layer another $SiO_2$ layer was grown by using atomic layer deposition to facilitate efficient binding of the myelin particles. In this work, a Kretschmann configuration based SPR sensor was used in wavelength modulation mode to study the binding kinetics of human IgM antibodies to central nervous system (CNS) myelin particles. Notably, myelin is mainly composed of lipid (70-85% of dry mass), protein (15-30% of dry mass), and has water content of about 40% [132]. Ryu et al. [133] reported the development of a novel and versatile membrane platform, which is continuous but spatially heterogeneous, to monitor the time-lapse dynamics of lipid-lipid interactions during the transient raft functional domain formation and sequential raft-associated receptor-ligand interactions. Regarding protein–protein interaction studies, detailed protocol for the protein immobilization on the sensor surface, analyte-binding analysis, affinity/kinetic measurements, as well as data analysis by using a BIAcore T200 system is described by Douzi [134]. Furthermore, Kim et al. [135] used Kretschmann configuration based SPR spectroscopy to demonstrate the binding of poly(ethylene oxide)-b-poly(propylene oxide)-b-poly(ethylene oxide) or PEO–PPO–PEO triblock copolymer to the model cell plasma membrane composed of supported lipid bilayer (SLB). Such studies are particularly important for the proper understanding of the cell membrane protective mechanism of poloxamer 188 (P188) triblock copolymer that involves polymer-membrane interaction kinetics. Most recently, Belkilani et al. [129] demonstrated multiparametric surface plasmon resonance (MP-SPR) technique to investigate the interactions of mono-rhamnolipids (mono-RLs) with biomimetic membranes. Using this technique, the group successfully characterized the evolution of liposome architecture from monolayer to phospholipid bilayer induced by mono-RL interactions with a detection limit of 2 μg mL$^{-1}$. Fig. 5 shows the sensorgram measured during mono-RLs – liposome interaction, at the of 670 nm wavelength. Notably, several other reviews on the use of SPR as a method for the study of interactions involving proteins and lipids can be found in the literature [97,136,137].

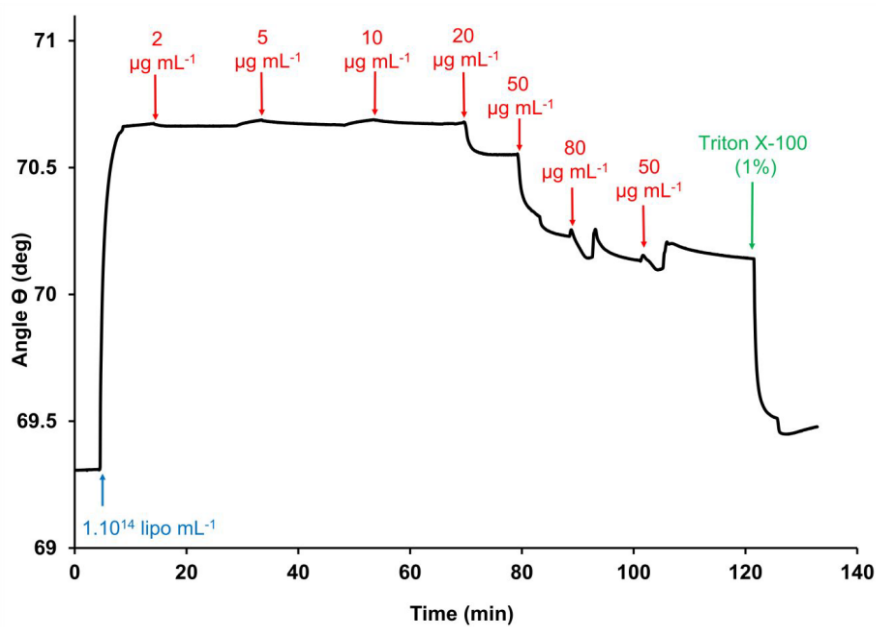

Fig. 5. SPR angle versus time sensorgram, measured at the of 670 nm wavelength, during the interaction of mono-RLs with liposomes. The figure is reproduced with permission from Belkilani et al. [129], Copyright © 2021 American Chemical Society

## 4.2 Detection of biomarkers

Biomarkers give vital information about the progression or regression of a disease and therefore have high clinical significance. Proper screening of different biomarkers, e.g., nucleic acids, hormones, proteins, and other biomolecules, is essential for the detection, diagnosis, and treatment of a disease to prevent its further spread, thereby significantly reducing the death rate. A number of strategies, including enzyme-linked immunosorbent assay (ELISA), polymerase chain reaction (PCR), immunofluorescence, western blotting, etc., have been already developed for the detection of disease biomarkers. Although efficient, these techniques are mainly limited by their operational complexity, the requirement of large sample volume, and longer processing time [138-140]. Recently, SPR based sensors have attracted much attention due to their simplicity, cost-effectiveness and capability to offer highly sensitive, and biomarker specific real-time measurement even with low sample concentration. This section intends to cover some of the recent applications of Kretschmann configuration-based SPR sensors for biomarker detection.

While the standard clinical protocols for evaluating cancer typically rely on tissue biopsy, which involves painful strategies using special needles and surgeries, SPR biosensors are emerging as a promising alternative capable of offering non-invasive liquid biopsy [141]. SPR sensors based on Kretschmann configurations have been exploited by many researchers for the analysis of cancer related biomolecules [142]. With its robustness and versatility, SPR sensors perform analysis in a broad range including plasma, serum, and saliva [143]. Detection of baculoviral inhibitor of apoptosis containing protein biomarker BIRC4 in dog-serum has also been reported [95]. Moreover, SPR assays are used for the detection of protein biomarkers in human sera in cases of pancreatic cancer [144,145], ovarian cancer [146], and coronary artery disease [147-149]. For early detection of cancer, SPR is used to detect cancer biomarkers through the development of point-of-care immunosensors. In one of such works, Uludag et al. [150] successfully detected prostate cancer total prostate-specific antigen (tPSA) with a detection limit of 0.29 ng mL$^{-1}$ (8.5 pM) by using PSA antibody-modified Au nanoparticles of size 40 nm. Recently, a high precision SPR (HP-SPR-3D) system is reported for the phenotypic screening of anti-cancer drugs using cancer cell lines [151].

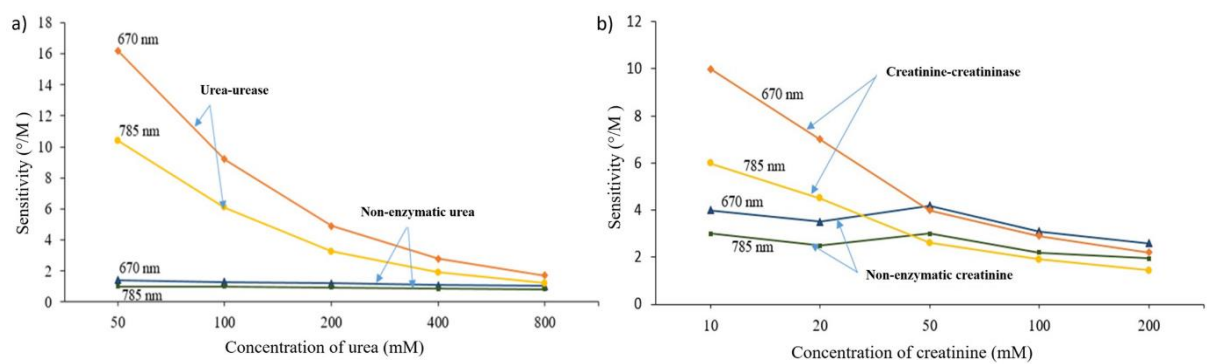

Fig. 6. Sensitivity of 50 nm-thick nano-laminated gold film vs concentrations of (a) urea-urease and (b) creatinine-creatininase samples at 670 and 785 nm optical wavelengths. The figure is reproduced from Menon et al. [152], Copyright: © 2018 Menon et al. Creative Commons Attribution License.

The kidney is one of the most vital organs of the human body that removes waste products from our body and maintains the balance of fluids and other chemicals. Alarmingly, kidney failure, which affects 10% of the worldwide population, is one of the leading causes of millions of deaths every year. Notably, the functioning of the kidney is generally monitored by observing the levels of urea and creatinine biomarkers [152,153]. In this context, SPR biosensors have shown enormous potential for application in such biomarker detections. Recently, a nano-laminated gold film (50-nm thick) based SPR sensor in Kretschmann configuration was used to detect urea and creatinine at various concentrations [152]. The group employed finite difference time domain (FDTD) simulation to determine the thickness of the nano-laminated gold film (that gives the best reflectivity), incident wavelength, and source angle for obtaining optimum SPR coupling and reflectivity. As shown in Fig. 6, enhanced sensitivities were obtained for bothenzymatic urea-urease sample (16.2°/M at 670 nm and 10.4°/M at 785 nm) and creatinine-creatininase sample (10°/M at 670 nm compared to 6°/M at 785 nm). In another work, Said et al. [154] used graphene-coated copper-based SPR sensor in Kretschmann configuration for urea-detection. Notably, the primary role of graphene in this work was to act as the preventive layer for Cu oxidation and enhance SPR reflectance intensity. Due to the high adsorption efficiency of graphene and the cost-effectiveness of copper over gold and silver, such a bio-recognized engineered device promises itself as an efficient SPR sensor, especially for biomolecules with carbon-based ring structure. Moreover, biosensors utilizing graphene-$MoS_2$ under Kretschmann based SPR configuration have also shown promising results for urea-detection in gas chromatography and calorimetric, fluorimetric and their respective analysis [155,156]. In the case of creatinine detection, Pothipor et al. [153] demonstrated an electrochemical (EC)-SPR instrumentation, with a detection limit of 0.19 μM, that used conducting-polymer based SPR sensor chip made of polypyrrole (PPy). The work involved the electropolymerization of a PPy film on the gold-coated SPR chip and subsequently monitoring the aggregation of starch-stabilized AgNPs induced by creatinine. The selectivity and specificity of the technique was ensured by investigating the effect of other substances, e.g., uric acid, albumin, etc., with Ag nanoparticles.

On the other hand, the excess of glucose in the blood causes diabetes. With a large population suffering from insulin resistance or insufficient insulin production by the body, the need for blood glucose levels monitoring has become necessary. The most conventional technique is the fringe prick blood test method which involves sampling blood through test strips and glucometer. Although effective, this method proves to be inconvenient for patients with self-monitoring glucose management. Propositions of Kretschmann configuration based SPR sensors for such analysis has already been done as an alternative. Recently, glucose detection by using Kretschmann configuration based SPR approach in angular modulation mode was demonstrated by Menon et al. [157]. The group utilized a nano-laminated (thickness: 50 nm) Au-Cr SPR sensor to evaluate the refractive indices of glucose levels at various concentrations. Another bimetallic Ag (30nm)/Au (17.5nm) thin film based SPR glucose sensor was reported by Mulyanti et al. [158] that showed sensitivity and figure of merit values of 4.7993°/RIU and 0.5173 at 633 nm incident wavelength for a glucose concentration of 0.08 g/dl. Most recently, other multi-layer, e.g., chromium/gold [159], Ag/Si3N4/Au [160], and SiO2 /gold/silver [161] thin film based SPR biosensors have shown excellent performances in the detection of glucose in blood or glucose/water solutions. Notably, different methods, including Taguchi method, FDTD simulation, COMSOL simulation, were used in such works to optimize the sensor design considerations.

The next major concerning malady is strokes. It mainly occurs when a blood vessel that carries oxygen and nutrients to the brain bursts or gets blocked by a clot [162]. Early diagnosis and appropriate treatment are of paramount importance for the prevention of strokes [163]. Several researchers have reported the reliability of SPR biosensors for the recognition of biomarkers related to stroke and their underlying mechanism [164-168]. Recent work by Harpaz et al. [169] on SPR biosensors for detecting stroke biomarkers is one such notable example. The group used a novel antibody functionalized $SiO_2$ plasmonic chip to detect stroke biomarkers n-terminal pro brain natriuretic peptide (NT-proBNP) and S100β. With a limit-of-detection less than 1 ng/mL, the reported sensor was found to be highly efficient for diagnosing strokes.

Impressively, SPR based biosensors are also used for the diagnosis of neurodegenerative diseases such as Alzheimer's disease (AD). Amyloid beta are peptides that form polymorphic fibrils and are active participants of diseases like AD and thus prove to be very important biomarker for AD. For instance, Ryu et al. [170] demonstrated the capability of SPR biosensor in the label-free monitoring of Alzheimer's *β*-amyloid (1- 42) fibrillation from the initial monomeric conditions on a solid surface. Recently a review article concerned with the applications of SPR-based biosensors have highlighted the detection of different biomarkers, e.g., aggregated *β*-amyloid and tau proteins, related to AD [171].

**4.3 Viral diagnostics**

With the first discovery of the yellow fever virus in 1901, at present, approximately 219 virus species are known to infect humans. There have been a complicated range of highly infectious viruses, e.g., influenza, dengue, human immunodeficiency, swine flu, Ebola, severe acute respiratory syndrome coronavirus (SARS- COV), SARS-COV-2 (COVID 19), etc., that are capable of producing severe health issues. They have special binding enzymes that bind to the host cells through pathogenic process, weakening the immune system. With unprecedented challenges emerging from such new viruses, effective global surveillance mechanisms need a boost in the subject [172].

Remarkably, among many other optical and/or non-optical techniques, SPR has become one of the most widely used optical techniques for viral diagnostics [173-185]. In a comparative study among SPR, resonant waveguide grating biosensing, and ELISA in the immunoassay evaluation of dengue virus, Hu et al. [178] found that Kretschmann configuration based SPR sensor is capable of providing accurate association ($k_{ass}$) and dissociation ($k_{diss}$) rates for the interaction between dengue virus (DENV) non-structural 1 (NS1) monoclonal antibodies (mAbs) with immobilized DENV NS1 proteins (a serological marker used for the early detection of the disease). The group further studied the affinity of antigen-antibody interactions. Interestingly, in a recent work, highly sensitive detection of DENV-2 E-proteins was demonstrated by performing SPR measurements at concentrations as low as 0.08 pM of the proteins (Fig. 7) [177]. For this work, an SPR sensor thin film of Au/dithiobis succinimidyl undecanoate/reduced grapheneoxide-polyamidoamine dendrimer/monoclonal antibody IgM (Au/DSU/NH$_2$rGO-PAMAM/IgM) was fabricated that offered excellent selectivity and enhanced penetration depth of evanescent waves for efficient SPR coupling. Moreover, Palau et al. [182] demonstrated the capability of SPR sensors in the real-time monitoring of interaction of RNAs of Hepatitis C virus (HCV) genome by immobilizing 5BSL3.2 on a sensor chip. In contrast to previously reported results, the group successfully demonstrated the complex formation of 5BSL3.2 with stem-loop SL2.

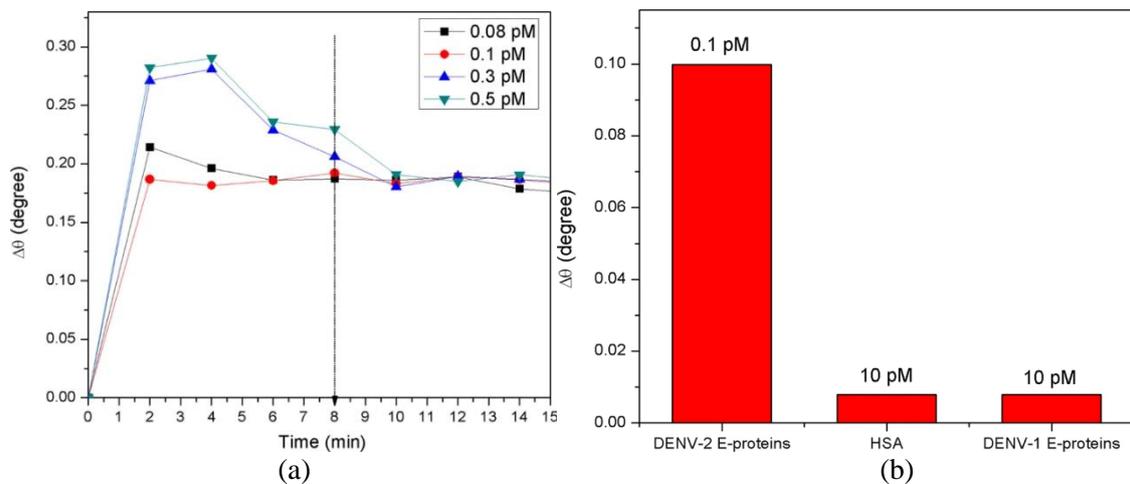

Fig. 7. (a) Sensorgram obtained during the real-time detection of DENV-2 E-proteins in contact with Au/DSU/NH$_2$rGO-PAMAM/IgM sensor film at different concentrations; (b) Selectivity of Au/DSU/NH$_2$rGO-PAMAM/IgM sensor for different target antigens. The figure is reproduced from Omar et al. [177], Copyright © The Author(s) 2020, Creative Commons Attribution 4.0 International License.

Regarding coronavirus infections, although the reverse transcriptase quantitative polymerase chain reaction (RT-qPCR) is considered the gold standard for detecting coronaviruses, it is time-consuming, requires a sophisticated testing facility, and sometimes give false results. In addition, while other sensing mechanisms, e.g., computed tomography (CT) scan and FET-based sensor topology using nasopharyngeal swabs, are still being explored [186,187], the need for real-time, ultra-sensitive technique could be fulfilled by SPR based sensors. In this context, in a pioneering work published in 2005, Chen et al. [183] used Biacore 3000 instrument to study the interaction of nonphosphorylated/phosphorylated nucleoprotein (N protein) of avian infectious bronchitis virus (IBV), which is a type of coronavirus, with a variety of target RNAs. Recently, Kretschmann configuration based SPR sensors was studied for real-time detection of severe acute respiratory syndrome coronavirus 2 (SARS-COV-2) [176]. Integrated layers of Si and BaTiO3 on Ag-metal layer with thiol tethered DNA as receptor has been proposed for sensing. The work was based on numerical analysis performed by using transfer matrix theory FDTD technique to characterize the performance parameters of the proposed sensor architecture. Moreover, another study, based on simulation, showed that a Kretschmann configuration based SPR sensor can detect 3.615 nm of virus concentration [188]. Furthermore, significant advances have also taken place for tackling pandemic by studying receptor recognition mechanisms of the coronavirus and angiotensin-converting enzyme 2 (ACE2). Through the SPR technique (Biacore 2000 instrument), bindings kinetics of the receptor binding domains (RBDs) of SARS-CoV, SARS-CoV-2, and chimeric SARS-CoV-2 with ACE-2 receptors were analysed [184]. By understanding of the receptor recognition mechanisms of the viruses, one can regulate their host range and infectivity. Importantly, it has been reported that ACE-2 is the receptor in humans that is recognized by both SARS-COV and SARS-COV-2 [189,190]. In a recent work by Das et al. [191], a sandwich-type SPR immunoassay model based on Au nanorods has been proposed in order to amplify the output signal. The group showed that evanescent field enhancement of 376.33% can be achieved for a 10 nm diameter Au nanorod coupled SPR sensor compared to the traditional Au nanosheet based sensors.

Table 2. Representative list of the biosensing applications of Kretschmann configuration based SPR sensors using different measurement methodologies

| Measurement methodology | Subject of study | Sensor chip used | Performance parameter(s) | References |
|---|---|---|---|---|
| Angular modulation | Serum YKL-40 estimation | CM5 sensor chip | LOD: 0.33 ng/ml | [192] |
| | Acetylcholine binding protein (AChBP) screening | CM5 and NTA sensor chips | S/N: 36 | [193] |
| | Detection of apple stem pitting virus (ASPV) coat protein (PSA-H, MT32) | Aptamer-modified sensor chips | LOD: 250 nm | [194] |
| | Detection of 2,4-dichlorophenol, a type of dioxin precursor | Anti-(2,4-dichlorophenol) antibody modified sensor chip | LOD: 20 ppb | [195] |
| | Detection of cytokeratin 19 fragment (CYFRA21-1), a biomarker related to lung cancer | Carboxyl-$MoS_2$-based chip | LOD: 0.05 pg/ml | [196] |
| | Detection of breast cancer cell line MCF-7 | Aptamer-modified gold chip | LOD: 136 cell/ml | [197] |
| Wavelength modulation | Detection of soluble vascular endothelial growth factor receptor (sVEGFR-1), a plasmatic protein marker of myelodysplastic syndromes (MDS) | Vascular endothelial growth factor (VEGF-A) immobilized gold coated replaceable sensor chip | LOD: 25 ng/ml | [198] |
| | Detection of tau-Aβ biomarker of Alzheimer disease | Mixed self-assembled monolayer (SAM) modified chip | LOD: 1 pM | [199] |
| | Measurement of specific antigen-antibody binding | Mouse IgG and goat IgG integrated gold chip | Dynamic range: $7.67 \times 10^{-3}$ RIU Resolution: $1.89 \times 10^{-6}$ RIU | [200] |
| | IgG antigen-antibody interaction | Rabbit IgG immobilized chip | RI resolution: $1.27 \times 10^{-6}$ RIU; Dynamic range: $4.63 \times 10^{-2}$ RIU | [201] |
| Intensity modulation | Avian influenza A H7N9 virus | H7-mAb immobilized Ag/Au (35/10 nm) chip | LOD: 144 copies/mL | [202] |
| | Nine common respiratory viruses - influenza A and influenza B, H1N1, respiratory syncytial virus (RSV), parainfluenza virus 1-3 (PIV1, 2, 3), adenovirus, and severe acute respiratory syndrome coronavirus (SARS) | Virus-specific oligonucleotide immobilized gold chip | LOD: Influ A—5 nM Influ B—1 nM PIV1—1 nM PIV2—2.5 nM PIV3—3.5 nM RSV—3 nM ADV—0.5 nM SARS—2 nM H1N1—3 nM | [203] |
| | Detection of lung cancer biomarkers EGFR (epidermal growth factor receptor) and PD-L1 (programmed death-ligand 1) | Antibody attached gold biochip | Sensitivity: $9.258 \times 10^3$%/RIU; Resolution: $8.311 \times 10^{-6}$ RIU | [204] |
| Phase modulation | Bovine serum albumin (BSA) antibody–antigen interaction | BSA immobilized gold chip | LOD: 0.5 ng/ml | [205] |
| | BSA/anti-BSA binding interaction | Protein immobilized gold sensor surface | Resolution: $8.8 \times 10^{-7}$ RIU; LOD: $7.7 \times 10^{-4}$ mg/ml | [206] |
| | Interaction of streptavidin–BSA complex | Biotin–protein immobilized gold chip | Sensitivity: 1.3 nM | [207] |
| | Rabbit IgG–goat–anti-rabbit IgG interaction | Rabbit IgG spotted microarray sensing chip | Phase resolution: 0.2°; RI resolution: $3 \times 10^{-5}$ RIU | [208] |
| Polarization modulation | Detection of single-stranded DNA (ssDNA) hybridization | Graphene/gold nanoparticle-based sensor chip | Dynamic range: $10^{-15}$ to $10^{-7}$ M; LOD: 500 aM | [209] |
| | Detection of lysozyme/antibody interaction | Lysozyme immobilized CM5 chip | LOD: $10^{-6}$ RIU | [210] |

| | Detection of influenza (H3N2) virus protein-antibody and DNA-DNA interaction | H3N2 virus protein/DNA immobilized gold surface | Sensor resolution: $4.36\times10^{-7}$ RIU; Detection limit: 8.6 nM (320 ng/mL), in the case of virus antibodies | [211] |

LOD: limit of detection or detection limit; S/N: signal to noise ratio; RI: refractive index; RIU: refractive index unit

## 5. Conclusion

The last few years have witnessed spectacular advances and facilitation in molecular biochemistry, materials science, and optoelectronics leading to the invention of several efficient laboratory-based and commercial SPR instruments [212]. Many commercially available SPR instruments, including SPRAutolab ESPRIT, Biocore series, Multiskop, and NanoSPR, have their principal based on prism coupling design. Notably, prism-coupled Kretschmann configuration is the most widely used configuration among different coupling schemes used in such SPR instrumentations. This chapter reviews the basic principles, instrumentation, and the most recent applications of Kretschmann configuration-based SPR biosensors.

SPR biosensors have now gained a promising and accomplished stature. Apart from their advanced analysing capability, their continuous progress and evolution towards the development of a highly accurate, sensitive, and user-friendly biosensor, they have been manifested as one of the most multifaceted devices for applications in diverse fields of molecular biology and biotechnology, particularly for the monitoring of dynamics and affinity of biomolecular interaction. The future of this technique lies entirely in the research and development of strategies related to antifouling, miniaturization, and initiations regarding the designing of innovative chemical chips. The real challenge which the field faces is its high-cost platforms and components. Once scientists and engineers develop a low-cost, highly-performance SPR sensor platform, SPR will be an irreplaceable tool for biological analysis and diagnosis.

**Acknowledgement**

AG acknowledges the University Grants Commission (UGC), India (Grant No. F.5-376/2014-15/MRP/NERO/2181), and Assam Science Technology and Environment Council, India (Grant No.: ASTEC/S&T/1614/8/2018-19/1159), for their support to the biophotonics research projects at JBC.